\documentclass[aps,prl,10pt,twocolumn,amsmath,amssymb,longbibliography,superscriptaddress,nofootinbib]{revtex4-2}
\usepackage{amsmath,amssymb,amsfonts} 	
\usepackage{graphicx}
\usepackage{hhline}
\usepackage[latin1]{inputenc}
\usepackage{subfigure}
\usepackage[normalem]{ulem}
\usepackage{xcolor}
\begin{document}

\title{Unified Treatment of Magnons and Excitons in Monolayer CrI$_3$ from Many-Body Perturbation Theory}

\author{Thomas Olsen}
\email{tolsen@fysik.dtu.dk}
\affiliation{CAMD, Department of Physics, Technical University of Denmark, 2800 Kgs. Lyngby Denmark}

\begin{abstract}
We present first principles calculations of the two-particle excitation spectrum of CrI$_3$ using many-body perturbation theory including spin-orbit coupling. Specifically, we solve the Bethe-Salpeter equation, which is equivalent to summing up all ladder diagrams with static screening  and it is shown that excitons as well as magnons can be extracted seamlessly from the calculations. The resulting optical absorption spectrum as well as the magnon dispersion agree very well with recent measurements and we extract the amplitude for optical excitation of magnons resulting from spin-orbit interactions. Importantly, the results do not rely on any assumptions on the microscopic magnetic interactions such as Dzyaloshinskii-Moriya (DM), Kitaev or biquadratic interactions and we obtain a model independent estimate of the gap between acoustic and optical magnons of 0.3 meV. In addition, we resolve the magnon wavefunction in terms of band transitions and show that the magnon carries a spin that is significantly smaller than $\hbar$. This highlights the importance of terms that do not commute with $S^z$ in any Heisenberg model description. 
\end{abstract}
\maketitle

Monolayers of CrI$_3$ comprise the first explicit realisation of two-dimensional (2D) ferrromagnetism \cite{Huang2017a} and have become a subject of extensive experimental as well as theoretical scrutiny in recent years \cite{Lado2017,Song2018,Zheng2018,Torelli2019,Sivadas2018,Huang2018,Klein2018a,Sun2019,Thiel2019,Huang2020,Lee2020,Zhang2020,Pizzochero2020,Qiu2021,Cenker2021}. The fundamental magnetic excitations - the magnons - are of fundamental interest in magnetic materials and the magnon dispersion can be used to extract the basic exchange interactions that govern the magnetic structure. In bulk CrI$_3$ the magnon dispersion has been determined by inelastic neutron scattering and shown to exhibit a gap of 4 meV at the Brillouin zone corner \cite{Chen2018a}, which was attributed to strong next-nearest neighbour Dzyaloshinskii-Moriya interactions. On the other hand, ferromagnetic resonance experiments has attributed the spinwave gap at the Brillouin zone corner to large Kitaev interactions that far exceeds the magnitude of nearest neighbor isotropic exchange \cite{Lee2020}. Both results, however, are in strong disagreement with first principles calculations of exchange, Kitaev and DM interactions for monolayer CrI$_3$ \cite{Xu2018,Xu2020} and the origin of the gap and as well as the nature and magnitude of the fundamental interactions seem to be unresolved at the moment. It should be emphasized though, that both experimental and theoretical estimates of magnetic interactions are based on anisotropic Heisenberg models (possibly including fourth order spin interactions) and the extracted parameters and interpretation of results may be strongly dependent on the underlying assumptions of the model. In the case of 2D CrI$_3$, the spinwave dispersion remains elusive since it cannot be obtained by inelastic neutron scattering. Nonetheless, magnetic Raman spectroscopy has recently shown that the energy of acoustic and optical magnons at the zone center for monolayers and bilayers agree very well with those of bulk \cite{Cenker2021}, which is corroborated by predictions of small interlayer exchange in bulk CrI$_3$ \cite{Sivadas2018,Soriano2019,Jiang2018c}. 

Most theoretical efforts to compute the magnon dispersion of CrI$_3$ is based on calculations within density functional theory (DFT), which is mapped to a Heisenberg-type model by means of either an energy mapping approach\cite{Xiang2013,Lado2017,Torelli2019,Xu2020} or the magnetic force theorem \cite{Kashin2020}. The predicted Heisenberg parameters may vary by a factor of two depending on the applied functional and model assumptions \cite{olsen_mrs}, but so far no DFT calculations have predicted the observed large gap at the Brillouin zone corner. One exception is a time-dependent density functional theory (TDDFT) study of bulk CrI$_3$ \cite{Ke2021}, where electronic correlation effects was shown to open a gap of 1.8 meV, even in the absence of spin-orbit coupling. Such studies are of crucial importance for the theoretical investigation of magnetic excitations since it is currently far from clear if it is the {\it ab initio} methods or the model assumptions that yield the large deviations from experiments.

From a theoretical point of view the magnons can be regarded as a coherent superposition of electron-hole excitations. Other flavors of such excitations include excitons and plasmons and all two-particle excitations can thus be obtained by diagonalizing a two-particle Hamiltonian that contains the appropriate interactions. In fact, the optical properties in 2D insulators are typically completely dominated by excitonic effects due to quantum confinement and reduced non-local screening in 2D \cite{Huser2013b,Cudazzo2015}. For example, in the case of MoS$_2$ the low energy absorption spectrum is dominated by a double excitonic peak that originates from the spin-orbit split valence band at the Brillouin zone corner \cite{Mak2010}. Recently, the optical absorption spectrum of CrI$_3$ was calculated using many-body perturbation theory and the results was shown to be in good agreement with experimental measurements of differential reflection \cite{Wu2019}.

In this letter we present first principles calculations of the generalized susceptibility in the framework of many-body perturbation theory including spinorbit coupling. The approximation for the susceptibility involves the summation of all ladder diagrams with static screened interactions, which amounts to solving the Bethe-Salpeter equation (BSE) \cite{Onida2002}. The poles include magnons as well as excitons and we extract the optical excitation spectrum as well as the magnon dispersion from the calculations. The spinorbit coupling is shown to open a gap of 0.3 meV at the Brillouin zone corner and gives rise to a finite amplitude for optical excitation of the magnons. Finally, by analyzing the magnon wavefunctions in terms of single-particle excitations we show that the strong spin-orbit coupling in the material results in a significant contribution of minority to majority spin excitations and the magnons thus carry angular momenta of $\sim-0.7\hbar$.

We consider the retarded dynamic four-component susceptibility in the Lehman representation which is written as
\begin{align}
    \chi^{\mu\nu}(\mathbf{r},\mathbf{r}',\omega) = \lim_{\eta \rightarrow 0^+} \sum_{\lambda \neq \lambda_0}\bigg( &\frac{n^\mu_{0\lambda}(\mathbf{r}) n^\nu_{\lambda0}(\mathbf{r}')}{\hbar \omega - (E_{\lambda}-E_{0}) + i\eta}
    \notag \\
    -& \frac{n^\nu_{0\lambda}(\mathbf{r}') n^\mu_{\lambda0}(\mathbf{r})}{\hbar \omega + (E_{\lambda}-E_{0}) + i\eta}\bigg),
    \label{eq:lehmann}
\end{align}
where $E_\lambda$ are eigenenergies, $\lambda_0$ denotes the ground state with energy $E_0$ and $n_{0\lambda^\mu}(\mathbf{r})=\langle0|\hat{n}^{\mu}(\mathbf{r})|\lambda\rangle$. Here the density operators are given by
\begin{align}
    \hat{n}^{\mu}(\mathbf{r}) = \sum_{s, s'} \hat\psi^\dag_{s}(\mathbf{r})\sigma^\mu_{s s'}\hat\psi_{s'}(\mathbf{r}),
    \label{eq:n_mu}
\end{align}
with $\mu,\nu\in[0,x,y,z]$ and $\sigma^\mu$ are the Pauli matrices augmented by the identity matrix $\sigma^0$. We refer to Ref. \cite{Skovhus2021} for details on the notation and formalism. It is convenient to express the susceptibility in a four-point spin basis such that $\chi^{\mu\nu}=\sigma^\mu_{s_1s_2}\sigma^\nu_{s_3s_4}\chi_{s_1s_2s_3s_4}$, since the four-point generalization of the susceptibility in Eq. \eqref{eq:lehmann} can then be shown to satisfy a Dyson equation within the Bethe-Salpeter approximation. The problem is most easily solved in a basis of Bloch states where it can be mapped to an eigenvalue problem
\begin{align}
    H(\mathbf{q})|\lambda(\mathbf{q})\rangle=E_\lambda(\mathbf{q})|\lambda(\mathbf{q})\rangle,
\end{align}
with the Hamiltonian given by
\begin{align}\label{eq:H}
H_{\substack{\mathbf{k}_2n_1n_2\\\mathbf{k}_4n_3n_4}}(\mathbf{q})=&\delta_{\mathbf{k}_2\mathbf{k}_4}\delta_{n_1n_3}\delta_{n_2n_4}(\varepsilon_{\mathbf{k}_2+\mathbf{q}n_1}-\varepsilon_{\mathbf{k}_2n_2})\\
-&(f_{\mathbf{k}_2+\mathbf{q}n_1}-f_{\mathbf{k}_2n_2})K_{\mathbf{k}_2n_2\mathbf{k}_2+\mathbf{q}n_1\mathbf{k}_4+\mathbf{q}n_3\mathbf{k}_4n_4}.\notag
\end{align}
Here $\varepsilon_{\mathbf{k}n}$ and $f_{\mathbf{k}n}$ are single-particle eigenenergies and occupation numbers respectively. The kernel $K$ contains the bare Coulomb interaction $v_\mathrm{c}$ and the (static) screened interaction $W$. The spin structure in a (hole-electron) basis of  $|\uparrow\uparrow\rangle,|\uparrow\downarrow\rangle,|\downarrow\uparrow\rangle,|\downarrow\downarrow\rangle$ can be written as
\begin{align}
    K=\begin{bmatrix}
v_\mathrm{c} - W & 0 & 0 & v_\mathrm{c} \\
0 & 0 & -W & 0\\
0 & -W & 0 & 0\\
v_\mathrm{c} & 0 &  & v_\mathrm{c} -W
\end{bmatrix}
\end{align}
and in a collinear description with spins aligned along the $z$-axis the transverse magnetic response ($\mu,\nu\in [x,y]$) is then decoupled from the density-density and longitudinal magnetic response functions ($\mu,\nu\in [0,z]$). When spinorbit coupling is included magnetic excitations may acquire a finite weight in $\chi^{00}$ and vice verca.

In the Tamm-Dancoff approximation the eigenstates can be expanded in two-particle Bloch states such that
\begin{align}
    |\lambda(\mathbf{q})\rangle=\sum_{\mathbf{k}n_1n_2}A_{\mathbf{k}n_1n_2}^{\lambda}(\mathbf{q})c^\dag_{\mathbf{k+q}n_1}c_{\mathbf{k}n_2}|\mathrm{GS}\rangle,
\end{align}
where $|\mathrm{GS}\rangle$ is the ground state and the full susceptibility may then be written in a plane wave representation as
\begin{align}\label{eq:susc}
\chi_{s_1s_2s_3s_4}^{\mathbf{G}\mathbf{G}'}(\mathbf{q},\omega)=\frac{1}{V}\sum_{\lambda}\frac{B^{\lambda}_{s_1s_2}(\mathbf{q},\mathbf{G})C^{\lambda}_{s_3s_4}(\mathbf{q},\mathbf{G}')}{\omega-E_\lambda(\mathbf{q})+i\eta},
\end{align}
where $\mathbf{q}$ is in the first Brillouin zone,
\begin{align}
&B^{\lambda}_{s_1s_2}(\mathbf{q},\mathbf{G})=\sum_{\mathbf{k}n_1n_2}\rho_{\mathbf{k}n_2s_2}^{\mathbf{k}+\mathbf{q}n_1s_1}(\mathbf{G})A^\lambda_{\mathbf{k}n_1n_2},\\
&C^{\lambda}_{s_1s_2}(\mathbf{q},\mathbf{G})=\sum_{\mathbf{k}n_1n_2}\rho_{\mathbf{k}n_2s_2}^{\mathbf{k}+\mathbf{q}n_1s_1*}(\mathbf{G})A^{\lambda*}_{\mathbf{k}n_1n_2}(f_{\mathbf{k+q}n_1}-f_{\mathbf{k}n_2}),\notag
\end{align}
and $\rho_{\mathbf{k}n_2s_2}^{\mathbf{k}+\mathbf{q}n_1s_1}(\mathbf{G})$ is the plane wave representation of the pair density $e^{-i\mathbf{q}\cdot\mathbf{r}}\psi^*_{\mathbf{k}n_2s_2}(\mathbf{r})\psi_{\mathbf{k+q}n_1s_1}(\mathbf{r})$. 

\begin{figure}[bt]
    \centering
    \includegraphics[width=\linewidth]{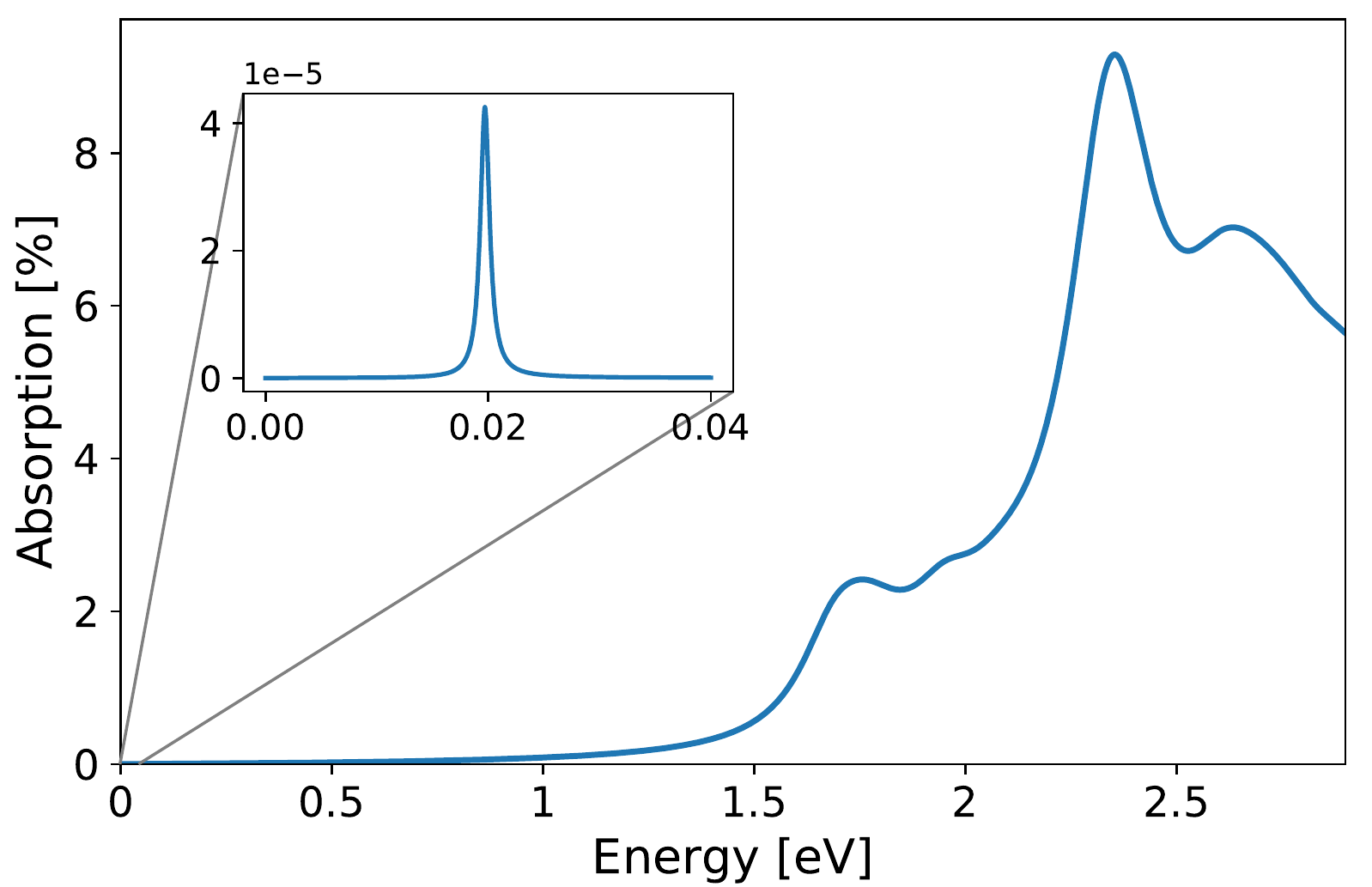}
    \caption{Absorption spectrum of monolayer CrI$_3$ calculated from the BSE. The inset show the low-energy part of the spectrum, which exhibits an optical magnon peak. The magnon peak has been shifted to its correct position by enforcing the acoustic magnon to have vanishing energy without spin-orbit coupling.}
    \label{fig:spectrum}
\end{figure}
For all our calculations we have used the experimental lattice constant of $a=6.867$ {\AA} and the ground state eigenenergies and orbitals were obtained with the local density approximation (LDA). The calculations were performed with the electronic structure package GPAW \cite{Enkovaara2010a,Larsen2017}, including spinorbit coupling \cite{Olsen2016a} and using a plane wave basis set and the projector-augmented wave method. We start by showing the optical absorption spectrum in Fig. \ref{fig:spectrum} obtained from
\begin{align}
    \mathrm{Abs}(\omega)=\mathrm{Re}[\sigma^\mathrm{2D}(\omega)]\bigg|\frac{2}{2+\sigma^\mathrm{2D}(\omega)}\bigg|^2,
\end{align}
where $\sigma^\mathrm{2D}$ is the 2D conductance (divided by $\varepsilon_0c$), which may be obtained directly from $\chi^{00}$. The spectrum was obtained using 50 eV plane wave cutoff, $24\times24$ $k$-point grid, 14 conduction electrons and 16 valence electrons for the Bloch state basis in the Bethe-Salpeter Hamiltonian. We included 162 conduction bands in the calculation of static screened interactions (same as Ref. \cite{Wu2019}) and a 2D Coulomb truncation scheme to decouple the screening from periodic images of the structure \cite{Rozzi2006}. When solving the Bethe-Salpeter equation we included a scissors shift of the conduction electrons to fix the band gap 2.59 eV as determined by GW calculations \cite{Wu2019}. The spectrum exhibits a first excitonic peak at 1.7 eV followed by a shoulder at 2.0 eV and then rise to a maximum at 2.3 eV. These features agree reasonable well with differential reflectrometry measurements on monolayer CrI3 \cite{Seyler2017}, which shows a shallow peak at 1.5 eV, a shoulder at 2.0 eV, and a maximum absorption at 2.2 eV. However, at higher energies there is some disagreement with the experimental spectrum, which does not decay above 2.3 eV but exhibits a second distinct peak at 2.7 eV. This is likely due to the fact that the quasi-particle band structure is not well represented by LDA with a scissors operator. In fact, scanning tunneling spectroscopy measurements of CrI$_3$ on graphite \cite{Qiu2021} have shown that the minority and majority conduction electrons are separated by a $\sim0.5$ eV gap whereas they overlap in LDA (see band structure below). We note that the binding energy of $\sim 1$ eV is in very good agreement with the analytical expression $E_\mathrm{b}=3/(4\pi\alpha^\mathrm{2D})$ \cite{Olsen2016} for 2D Mott-Wannier excitons ($\alpha^\mathrm{2D}$ is the 2D polarizability, which is 3.2 {\AA} for CrI$_3$) \cite{Haastrup2018}.

When analyzing the eigenvalue spectrum obtained from diagonalizing the Bethe-Salpeter Hamiltonian, one observes two low-lying eigenvalues that are well separated from the remaining part of the spectrum. Turning to $\chi^{+-}=\chi_{\uparrow\downarrow\downarrow\uparrow}$ we see that these states have weights that are close to 6 Bohr magnetons, which comprises the total integrated weight of $\chi^{+-}$. These states thus represent strong transverse magnetic excitions and we identify these as the acoustic and optical magnons - the latter of which has a finite optical weight (in $\chi^{00}$) due to spin-orbit coupling. We note that the assignment of acoustic and optical modes is supported by the fact that the optical mode has neglectable weight for the $\mathbf{G=0}$ component of the susceptibility \eqref{eq:susc} but a strong weight for any reciprocal lattice vector in a direction that connects the two Cr atoms. As show in the Supplementary Material, the magnons are in some respects much more demanding to converge than the excitonic spectrum and for the calculation described above these are located at negative energies. Moreover, it has previously been shown that even for converged results the present approach introduces a gap error that positions the acoustic magnon at a finite energy \cite{Muller2016}, whereas Goldstones theorem implies that its energy must be zero in the absence of spin-orbit coupling. Nevertheless, we have indicated the gap error corrected low-energy spectrum in Fig. \ref{fig:spectrum}, which shows the optical magnon, with an optical weight determined to be roughly 5 orders of magnitude lower that the first excitonic peak. This implies that it should, in principle, be possible to probe the optical magnons directly from infrared absorption or reflection spectroscopy, although decoupling the effect of phonons may be non-trivial.

\begin{figure}[tb]
    \centering
    \includegraphics[width=\linewidth]{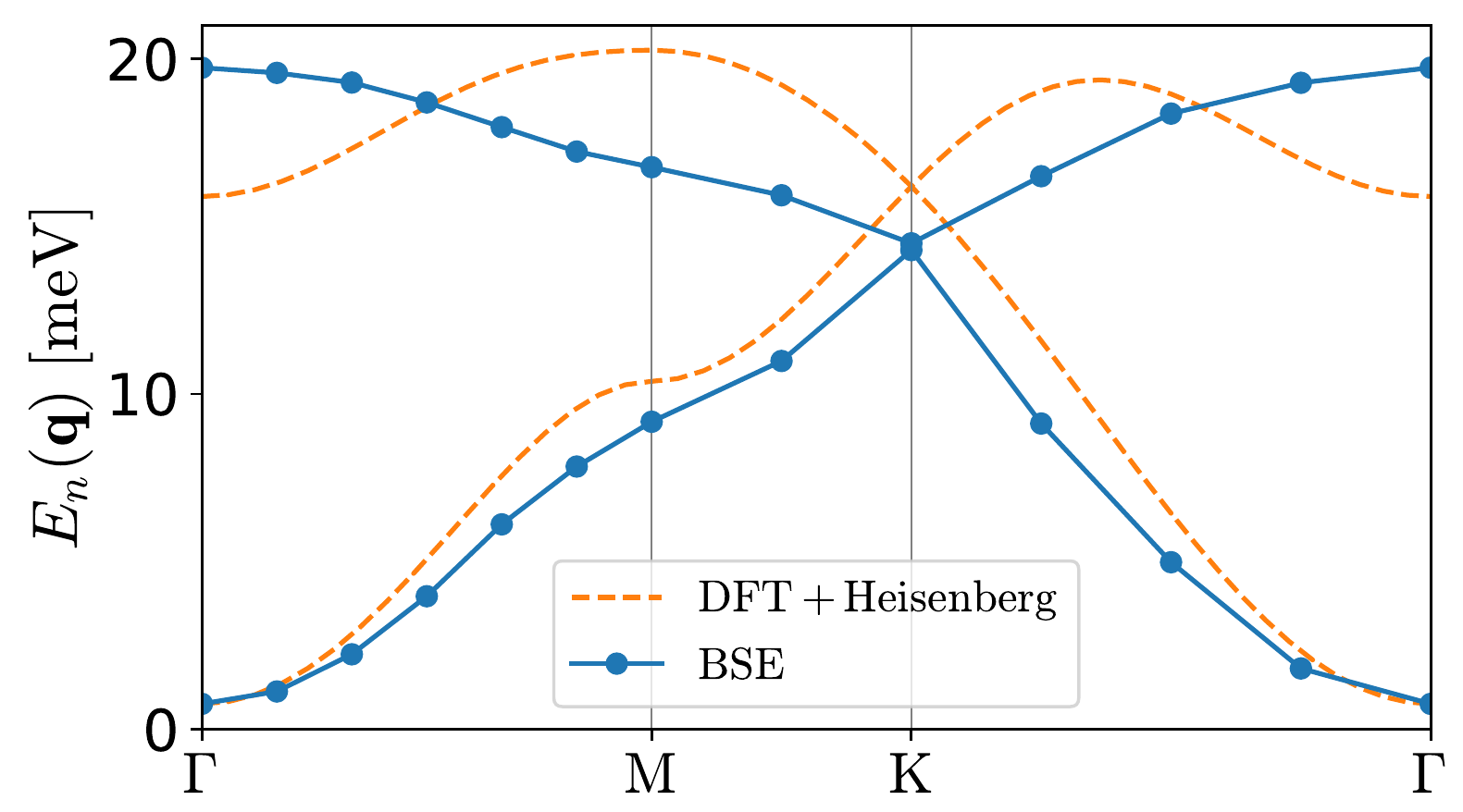}
    \caption{Dispersion relation for the acoustic and optical magnons modes along high symmetry paths in the Brillouin zone. The dashed line shows the spinwave dispersion obtained from a Heisenberg model with parameters obtained from DFT.}
    \label{fig:dispersion}
\end{figure}
In order to investigate the magnon spectrum further we have diagonalized the BSE Hamiltonian for wavevectors $\mathbf{q}$ along high-symmetry paths in the Brillouin zone. The calculations was performed using a plane wave cutoff of 150 eV, a $12\times12$ regular $k$-point grid, 300 bands in the screened interaction, 42 valence bands and 14 conduction bands in the diagonalization of the BSE Hamiltonian. This provides a rough magnon dispersion that is not well converged. In order to obtain a converged dispersion we have performed extrapolation of screening bands, plane wave cutoff and conduction bands to the infinite basis limit for each $\mathbf{q}$-point along the path (See Supplementay Material for details). The result is shown in Fig. \ref{fig:dispersion} and we note that the magnon energy of 19.7 meV for the optical mode at the $\Gamma$-point is rather  close to the experimentally measured value of 17 meV for bilayer CrI$_3$ \cite{Cenker2021}. We also display the spinwave dispersion obtained from a Heisenberg model with diagonal exchange interaction of the form 
\begin{align}\label{eq:model}
H=-\frac{1}{2}\sum_{\langle ij\rangle}J_{ij}\mathbf{S}_i\cdot\mathbf{S}_j-\frac{1}{2}\sum_{\langle ij\rangle}\lambda_{ij}S_i^zS_j^z-A\sum_i(S_i^z)^2,
\end{align}
where the isotropic exchange $J_{ij}$, anisotropic exchange $B_{ij}$, and single ion anisotropy constants were determined from a total energy mapping analysis. We include up to third nearest neighbors and obtain $J_1=2.4$ meV $J_2=0.6$ meV and $J_3=-0.4$ meV, which is in good agreement with previous LDA calculations \cite{Xu2020}. The spinwave gap is determined from $A$ and $\lambda_{ij}$ to be 0.76 meV and we use this to pin the acoustic magnon at $\mathbf{q=0}$ in the BSE spetrum. The large deviations between the two dispersion relations can be regarded as a strong limitation of the model \eqref{eq:model}, which has previously been used extensively to model magnetism in 2D materials \cite{Lado2017,Torelli2019c,Torelli2020a}. In particular, the model assumes diagonal exchange interaction and do not include DM interactions, Kitaev interactions or higher order spin interactions. There has recently been several attempts to augment the model with various types of spin interactions \cite{Xu2018,Liu2018a,Xu2020,Kartsev2020}, but presently the role and magnitude of the different terms is not completely clear. In particular, there is strong disagreement between first principles calculations \cite{Xu2018,Xu2020} and experiments \cite{Lee2020} regarding the magnitude of Kitaev interactions. In both cases, however, higher order spin interactions \cite{Hoffmann2020} are not taken into account, which may have a significant influence on the conclusions.

\begin{figure}[tb]
    \centering
    \includegraphics[width=\linewidth]{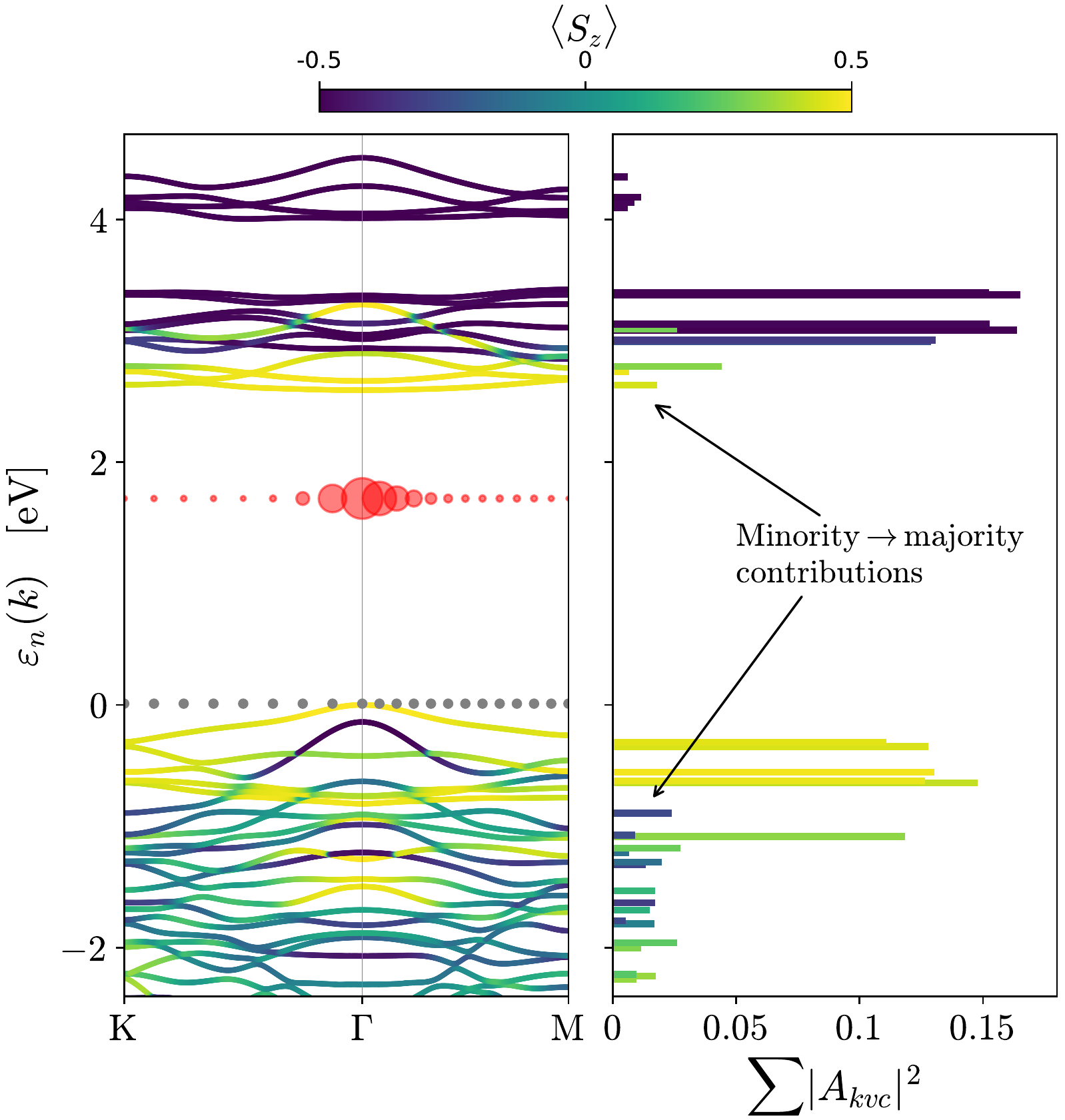}
    \caption{Left: LDA band structure of CrI$_3$ including a scissors operator that puts the band gap at the GW value of 2.59 eV. The dots show the $k$-space wavefunctions for the acoustic magnon (grey) and the lowest lying exciton (red). Right: band resolved wavefunction for the acoustic magnon (see text).}
    \label{fig:bands}
\end{figure}
A prominent feature of the BSE dispersion is the fact that it exhibits a gap between the acoustic and optical branches at $q=K$. We determine this gap to be 0.3 meV (see supplementary material), whereas the gap vanishes in the model \eqref{eq:model} by construction. In general, such a gap in honeycomb structures is often attributed to DM interactions \cite{Mook2014}, but since the nearest neighbor DM interaction vanishes by symmetry in CrI$_3$ a DM mediated gap opening must originate from second nearest interactions. The first principle value for the gap appears to be in line with DFT calculations of the second neighbor DM interaction, but is much smaller than the experimental gap of 4 meV in bulk CrI$_3$. Unfortunately, there are no experimental measurement of the gap in monolayer CrI$_3$, but it seems unlikely that this deviation (by an order of magnitude) can be attributed to the difference in electronic structure between bulk and monolayer CrI$_3$.

In order to gain more insight into the magnonic and excitonic excitations we show the wavefunctions of collective excitations in $k$-space ($\sum_{vc}|A_{kvc}|^2$) along with the band structure (including the scissors shift) in Fig. \ref{fig:bands}. The magnon wavefuncion has identical weights at all $k$-points indicating strong localization in real-space. This is in accordance with the Heisenberg model in which the magnon is a linear combination of electron-hole excitations (spin flips) at individual sites. In contrast, the excitonic wave function is strongle localized in $k$-space signifying an exciton of the Mott-Wannier type with a considerable delocalization in real space. In the absence of spinorbit coupling the magnons are solely composed of majority to minority band transitions. In the presence of strong spinorbit coupling it is thus of interest to resolve the magnon in terms of band transition in order to see to what extend that picture holds true. For a particular $k$-point we can consider the quantity $\sum_{v(c)}|A_{kvc}|$, which yields the weight originating from a particular conduction(valence) band summed over all valence(conduction) bands. This is shown in Fig. \ref{fig:bands} and we see that the most prominent contribution originate from the highest majority valence bands and lowest minority conduction bands as expected. There is, however, also a significant contribution from minority to majority transitions, which will inevitably render the magnon angular momentum less than $-\hbar$. To quantify this we calculate the expectation value $S_z$ of the magnon states, which is expressed as $\langle\lambda|S^z|\lambda\rangle=\sum_{kvv'cc'}A^*_{kvc}A_{kv'c'}(S^z_{cc'}-S^z_{vv'})$ with $S^z_{vv'(cc')}=\langle v(c)|S^z|v'(c')\rangle$. For both the acoustic and optical magnons at $q=\Gamma$ we obtain $\langle\lambda|S^z|\lambda\rangle=-0.7\hbar$ showing that the magnons indeed deviate from the classical picture of spin-1 particles. This emphasizes the importance of a spin model that does not commute with $S^z$, since for example the model \eqref{eq:model} can only yield excitations of integer spin and cannot be expected to describe the magnons properly.

In summary we have shown that both magnons and excitons in 2D CrI$_3$ are rather accurately captured as bound states when solving the Bethe-Salpeter equation and they can be distinguished by their respective weights in $\chi^{00}$ and $\chi^{+-}$. In contrast to common approaches based on spin models the calculated magnon dispersion does not rely any model assumptions regarding the nature of spin interactions and the dispersion thus comprises a highly useful starting point for unravelling the nature magnetic interactions. In particular, spinorbit coupling induces a gap between the acoustic and optical magnon branch of 0.3 meV, which could be as signature of next-nearest neighbor DM interactions. We note that a recent study based on a similar approach with a fitted model Hamiltonian \cite{Costa2020} yielded a gap of 2 meV, which is much closer to experimental value in bulk \cite{Chen2018a}. The reason for this difference is not clear at present. Moreover, the magnon wavefunction was shown to involve significant contributions from majority to minority transitions and the spin character of the magnons are reduced by 30 {\%} compared to the integer value obtained without spinorbit coupling. This highlight the crucial importance of terms in spin models that do not commute with $S_z$ and the presented magnon dispersion could perhaps be used as a benchmark when investigating the role of Kitaev, DM and higher order spin interactions in generalized Heisenberg models.

This work was supported by Villum Fonden, grant number 00029378.

%

\end{document}